\documentclass[aps,prd,showkeys,showpacs,amssymb,
amsfonts,epsf,preprintnumbers,nofootinbib,
superscriptaddress]{revtex4}

\usepackage[dvips]{graphicx}
\usepackage{bm,latexsym,amsmath,amssymb,amsfonts,color}

\def\be {\begin{equation}}
\def\ee {\end{equation}}
\def\ba {\begin{eqnarray}}
\def\ea {\end{eqnarray}}

\def\bi {\begin{itemize}}
\def\ei {\end{itemize}}
\newcommand\beq{\begin{eqnarray}}
\newcommand\eeq{\end{eqnarray}}
\newcommand{\bea}{\begin{eqnarray}}
\newcommand{\eea}{\end{eqnarray}}

\usepackage{graphicx}
\usepackage{dcolumn}
\usepackage{bm,morefloats,color}

\usepackage{amssymb,bm}

\def\X5sp{{\rm X}_5}
\def\Y3sp{{\rm Y}_3}
\def\Z3sp{{\rm Z}_3}

\begin{document}

\title{Braneworlds with field derivative coupling to the Einstein tensor}

\author{Masato Minamitsuji}
\affiliation{Multidisciplinary Center for Astrophysics (CENTRA), Instituto Superior T\'ecnico, Lisbon 1049-001, Portugal.}

\begin{abstract}
In this paper, we investigate the Randall-Sundrum type braneworld models in the scalar-tensor theory with field derivative coupling to the Einstein tensor. We first formulate the generalized junction conditions of the metric and the scalar field on the timelike codimension-one hypersurface (=brane). With the use of these junction conditions, we then derive the Minkowski and de Sitter brane solutions embedded into the $Z_2$-symmetric five-dimensional anti-de Sitter (AdS) bulk spacetime. The configuration of the scalar field depends on the slice of the AdS spacetime. These branes are supported by the tension and not coupled to the scalar field. The Minkowski brane solution can be obtained when the brane tension is tuned to the bulk contributions. Once this tuning relation is broken, the de Sitter brane solutions are obtained. The de Sitter brane solutions have two branches. One has the smooth limit to the case where the scalar field becomes trivial, while the other branch does not. The latter branch has the upper bound on the brane tension, where the expansion rate vanishes. Finally, we investigate the low energy effective gravitational theory realized on the brane, which is given by the four-dimensional Einstein-scalar theory with the corrections from the bulk.
\end{abstract}
\maketitle

\section{Introduction}

Recently,
the modified gravity theories have received the renewed interests (see e.g., \cite{mg} and references therein).
The main motivation arises due to the fact
that recent cosmological observational data have suggested 
the necessity of introducing a couple of the mysterious components in the past and present universe
to explain the cosmic history consistently. 
One of them is the present day's accelerated expansion of the universe.
Except for the simplest explanation by introducing a cosmological constant or an alternative dark energy source,
modifying the gravitational force law at the cosmological distance scales
provides another possible way of explaining this.
The other important issue is 
what causes the inflationary expansion of the primordial universe.
The higher curvature corrections to general relativity motivated by the unified theory,
such as the $R^2$ corrections \cite{sta},
have also been thought to be promising ways to providing the successful models of inflation.

On the other hand,
some kinds of the candidates of the unified theory and phenomenological extension of general relativity,
such as superstring/M-theory,
predict the higher dimensional spacetime instead of the four-dimensional one.
Although the idea of introducing an extra dimensional space has a long history,
the recent studies have been motivatived by the developments in superstring/M-theory.
In contrast to the traditional higher-dimensional approach
where the extra dimensions should be extremely small,
string theory has suggested a variety of ingredients in the extra dimensional space 
which may be useful for cosmological and phenomenological model buildings.
An interesting approach to the cosmological model construction
has been the so-called braneworld model \cite{add,add2,rs2,rs1,ds,ds2}
which assumes that our universe corresponds to a four-dimensional brane located
at the boundary of the (warped) extra dimensional space
and all the fundamental interactions except for gravity are confined to the brane
(see also e.g., \cite{bwg,bwg2,bwg3} and references therein).  
Such a possibility has motivated us to study the possible signatures
from extra dimensions which should be detected in the future experiments and observations.
Among the various models,
in this paper
we would like to investigate some extension of the so-called Randall-Sundrum type models
where our brane universe is located at the boundary of the $Z_2$ symmetric five-dimensional anti-de Sitter (AdS) 
bulk spacetime \cite{rs1,ds,ds2}.

The higher curvature corrections to the five-dimensional Randall-Sundrum braneworld models have been considered,
e.g.,
by adding the Gauss-Bonnet corrections to the bulk gravity action
(see e.g., \cite{gbbw,gbbw1,gbbw2,gbbw3,gbbw4})
or the induced gravity term to the brane action \cite{dgp1,dgp2}.
However, taking the fact that superstring/M-theory predicts ten- or eleven-dimensional spacetime into consideration,
the five-dimensional theory
may also be an effective description of the more fundamental higher-dimensional theory. 
In the reduced five dimensional theory,
the moduli fields parametrizing the geometry of the compactified internal space
are nonminimally coupled to the five-dimensional gravity.
The five-dimensional effective action is then given by a class of the scalar-tensor theories.
The braneworld models in the five-dimensional scalar-tensor theory have been considered in e.g., \cite{bs,bs2,bs3,bs4,bim,bim2,bim3}.
On the other hand,
it is known that in superstring theories
quantum corrections appear as the higher-rank curvature terms \cite{st,st2,st3,st4}
plausibly given by the Lovelock terms \cite{love}
which do not give the higher derivative terms in the equations of motion.
The general scheme of the dimensional reduction from the higher-dimensional Lovelock gravity has been discussed in \cite{van}, and the lower dimensional effective theory should be written in the context of
the most general scalar-tensor theories recently discussed in e.g., \cite{hor,nic,def,def2,def3,kyy,char}.
In addition to the self-interactions of the field derivatives,
the lower dimensional effective theory also contains nonminimal coupling 
of field derivatives to the curvature.
The modifications of the brane dynamics in the effective five-dimensional gravity 
would appear in the following two ways:
The first is via the modification of the solutions to the bulk gravity.
The second is via the generalization of the Israel junction condition \cite{israel}
(See e.g. \cite{sb,sb1,sb2,sb3} for the earlier works in this direction).

Among the various models with field derivative coupling,
in this paper we will especially focus on the field derivative coupling to the Einstein tensor $G^{AB}\partial_A \phi \partial_B \phi$
and investigate its roles in the construction of the warped braneworld models.
The highest derivative terms in the equations of motion 
are still of the second order
with thanks to the contracted Bianchi identities $\nabla_A G^{AB}=0$.
Thus among the various types of field derivative coupling to gravity,
the coupling to the Einstein tensor provides almost the unique possibility for getting the second order equations of motion.
Such a theory has attracted much interest 
from cosmology \cite{ame,cap,sch,sar,gra,gra2,cd1,cd2,cd3,ger,ger2,gw,sal,papa,qiu} and black hole \cite{cj,koly,koly2,rinaldi}
points of view.
In the context of the braneworld model,
the localization of the bulk fields with various spins onto the brane
via their nonminimal derivative coupling to curvature tensors was investigated in Ref. \cite{brane_ger}.

In this paper,
we first formulate the generalized junction conditions 
for a timelike, codimension-one singular hypersurface (=brane) embedded into the general  $D$-dimensional bulk spacetime.
The derivatives of the metric components and the scalar field with respect to the bulk coordinate
would have discontinuity across the brane.
The mathematical structure of the generalized junction conditions 
remains the same as in the case without the field derivative coupling,
except that they also contain the brane intrinsic quantities (See also \cite{Padilla:2012ze}).
With the use of the junction conditions,
we apply them for the construction of the braneworld models
embedded into the $Z_2$-symmetric five-dimensional  AdS spacetime.
We consider the Minkowski and de Sitter branes embedded into the 
five-dimensional AdS and supported by a tension.
Finally, we derive the low energy effective gravitational theory  realized on the brane.

The paper is constructed as follows:
In Sec. II, we derive the equations of motion in the given scalar-tensor theory with field derivative coupling
to the Einstein tensor.
In Sec. III, we derive the generalized metric and scalar field junction conditions
by integrating the equations of motion acrosss the brane.
In Sec. IV, we derive the Minkowski and de Sitter brane solutions embedded into the $Z_2$-symmetric five-dimensional 
AdS spacetime.
In Sec. V, we discuss the low energy effective gravitational theory on the brane.
The last Sec. VI is devoted to giving a brief summary.

\section{The equations of motion}

We consider the $D$-dimensional spacetime with the $(D-1)$-dimensional, 
timelike, codimension-one boundary hypersurface (=brane)
in the $D$-dimensional scalar-tensor theory
with the field derivative coupling to the Einstein tensor 
\bea
\label{action}
S&=&
\int_M d^D X\sqrt{-g}
\Big(
\frac{1}{2\kappa^2}R
-\frac{1}{2}
\big(g^{AB}-
z \kappa^{\frac{4}{D-2}} G^{AB}\big)
\partial_A \phi
\partial_B \phi
-V(\phi)
+{\cal L}_M[g,\Psi]
\Big)
\nonumber\\
&+&
\int_\Sigma d^{D-1}x \sqrt{-q}
{\cal L}_b[q,\phi,\psi],
\eea
where $g_{AB}$ represents the metric of the $D$-dimensional bulk spacetime,
$G_{AB}$ is the Einstein tensor associated with $g_{AB}$,
and $q_{\mu\nu}$ is the induced metric of the $(D-1)$-dimensional brane.
$\phi$ represents the bulk scalar field with mass dimension $\frac{D-2}{2}$, 
$V(\phi)$ does its bulk potential
which does not include the contribution from the brane.
$X^A$ and $x^{\mu}$ are the bulk and brane coordinates, respectively.
The constants $\kappa$ and $z$ 
represent the gravitational and nonminimal coupling constants, respectively,
where $\kappa$ has mass dimension $-\frac{D-2}{2}$
and $z$ is dimensionless. 
${\cal L}_M[g,\Psi]$ and ${\cal L}_b[q,\phi,\psi]$ represent the 
Lagrangian densities of the matter fields
living in the bulk  (other than the scalar field) $\Psi$
and those on the brane $\psi$, respectively.

The metric of the $D$-dimensional bulk spacetime $(M,g_{AB})$ is represented by the Gaussian normal coordinate
\bea
\label{GN}
ds^2= g_{AB}dX^A dX^B=dy^2+ q_{\mu\nu} (x^{\mu},y)dx^{\mu}dx^{\nu},
\eea
where $y$ denotes the coordinate along the $D$-th direction and 
$x^{\mu}$ does that of the $(D-1)$-dimensional spacetime along the brane.
The brane $(\Sigma, q_{\mu\nu})$, i.e., the boundary singular hypersurface labeled by the coordinate $x^{\mu}$,
is assumed to be located at $y=0$.
The extrinsic curvature tensor on each constant $y$ slice
is given by 
\bea
\label{extrinsic}
K_{\mu\nu}=\frac{1}{2}q_{\mu\nu,y},\quad
K=q^{\mu\nu}K_{\mu\nu}=\frac{1}{\sqrt{-q}}\partial_y \big(\sqrt{-q}\big).
\eea
The nonzero components of the curvature tensors and derivatives of the scalar field in the $D$-dimensional spacetime
Eq. (\ref{GN}) are listed in the Appendix A.

Varying the action \eqref{action} with respect to the metric $g_{AB}$,
the gravitational equation of motion is given by
\bea
\label{metric_eom}
G_{AB}
+z\kappa^{\frac{2D}{D-2}}  L_{AB}
=\kappa^{2} T_{AB}+\kappa^2T_{AB}^{(M)}
+ \kappa^2 S_{\mu\nu} q^{\mu}_A q^{\nu}_B\delta (y),
\eea
where 
\bea
T_{AB}&=&\nabla_A \phi \nabla_B \phi
     -\frac{1}{2}g_{AB}\nabla^C\phi \nabla_C\phi
   -V(\phi)g_{AB},
\nonumber\\
L_{AB}&=&
-\nabla_A \nabla_B\phi \Box \phi
 +\nabla_A \nabla_C\phi \nabla_B \nabla^C\phi
 +\nabla^C \phi\nabla^D\phi R_{ACBD}
-\frac{1}{2}\nabla_A\phi\nabla_B\phi R
\nonumber
\\
&+&2\nabla_{(A} \phi R_{B) C}\nabla^C\phi
-\frac{1}{2}G_{AB} \nabla_C\phi \nabla^C\phi
\nonumber\\
&+&g_{AB}
\Big(
-R^{CD}\nabla_C\phi \nabla_D\phi
+\frac{1}{2}
\big(\Box\phi\big)^2
-\frac{1}{2}
\nabla_C\nabla_D\phi
\nabla^C\nabla^D\phi
\Big),
\eea
and
\bea
T_{AB}^{(M)}:=-\frac{2}{\sqrt{-g}} 
 \frac{\delta}{\delta g^{AB}}\big(\sqrt{-g}{\cal L}_M\big),
\quad
S_{\mu\nu}:=-\frac{2}{\sqrt{-q}} 
 \frac{\delta}{\delta q^{\mu\nu}}\big(\sqrt{-q}{\cal L}_b\big),
\eea
represent the energy-momentum tensors of the bulk matter fields
(other than the scalar field) and that of the brane localized matter fields, respectively.

Varying the action \eqref{action}
with respect to the scalar field $\phi$,
the scalar field equation of motion is given by 
\bea
\label{scalar_eom}
\Big(g^{AB}-z\kappa^{\frac{4}{D-2}} G^{AB}\Big)\nabla_A \nabla_B\phi=
\frac{dV}{d\phi}
-\frac{\delta {\cal L}_b}{\delta\phi}\delta(y).
\eea

\section{The junction conditions}

After giving the equations of motion,
by integrating them across the brane with respect to $y$,
we derive the junction conditions of the metric components and the scalar field.

First, in order to define the unique induced metric and the amplitude of the scalar field on the brane,
we assume that the metric components in \eqref{GN}
and the scalar field amplitude $\phi$
are continuous across the brane
\bea
\label{cont}
\big[q_{\mu\nu}\big]^+_-=0,\quad
\big[\phi\big]^+_-=0,
\eea
where $[A]:=A|_{y=0+}-A|_{y=0-}$ denotes the discontinuity 
of the physical quantity $A$ across the brane at $y=0$.
We also note that the covariant derivatives of the scalar field with respect to the
intrisic metric $q_{\mu\nu}$, such as $\phi_{|\mu}$ and $\phi_{|\mu\nu}$,
as well as the intrinsic curvature terms, such as ${}^{(q)}R$,  are also continuous across the brane.
However,
the first order derivatives of the metric components and the scalar field with respect to 
the bulk coordinate $y$,
$K_{\mu\nu}$ and $\phi'$,
have the discontinuity across the brane.

The $(y,y)$ and $(y,\mu)$ components of the gravitational equation \eqref{metric_eom}
do not contain the second order derivatives with respect to $y$,
which provide the Dirac delta function $\delta(y)$.
On the other hand,
the $(\mu,\nu)$ component of the gravitational equation \eqref{metric_eom}
contains the second order derivatives such as 
\bea
\label{co1}
G_{\mu\nu}
&=&-K_{\mu\nu,y}+ q_{\mu\nu} K_{,y}+\cdots,
\eea
and
\bea
\label{co2}
L_{\mu\nu}
&=&
-\phi_{|\mu\nu}\phi''
+\phi_{|\mu}\phi_{|\nu} K_{,y}
-\phi^{|\alpha} \phi_{|\mu} K_{\nu\alpha,y}
-\phi^{|\alpha} \phi_{|\nu} K_{\mu\alpha,y}
+\frac{1}{2}\big(K_{\mu\nu,y} -q_{\mu\nu}K_{,y}\big)
     \phi^{|\rho}  \phi_{|\rho} 
\nonumber\\
&+&
q_{\mu\nu}\Big(K_{\rho\sigma,y}\phi^{|\rho}\phi^{|\sigma} +\phi'' {}^{(q)}\Box \phi\Big)
-\frac{1}{2}\partial_y
\Big\{
\big(
K_{\mu\nu}-q_{\mu\nu}K
\big)
(\phi')^2
\Big\}
+\cdots,
\eea 
where terms of $\cdots$ are composed of at most the first order derivatives with respect to $y$, 
$A_{|\mu}$ represents the covariant derivative of $A$ with respect to the $(D-1)$-dimensional metric $q_{\mu\nu}$,
and ${}^{(q)}\Box A:=q^{\rho\sigma}A_{|\rho\sigma}$ is the $(D-1)$-dimensional d'Alembertian.
We note that in this paper
a prime denotes the derivative with respect to $y$.

Similarly,
in the scalar field equation of motion,
the second order derivative terms with respect to $y$ are given by
\bea
\label{co3}
&&\big(g^{AB}-z\kappa^{\frac{4}{D-2}} G^{AB}\big)\nabla_{A}\nabla_B\phi
\nonumber\\
&=&
\Big(
1
+\frac{z\kappa^{\frac{4}{D-2}}} {2}    
 {}^{(q)}R
\Big)\phi''
+z\kappa^{\frac{4}{D-2}}
\Big\{
     \big(K_{\rho\sigma,y} -q_{\rho\sigma} K_{,y}\big)
\phi^{|\rho\sigma}
-\frac{1}{2}
\partial_y
\Big(
\big(K^2-K^{\rho\sigma}K_{\rho\sigma}\big)
\phi'
\Big)
\Big\}
+\cdots.
\eea

Substituting \eqref{co1} and \eqref{co2} into the $(\mu,\nu)$ component of
the gravitational equation of motion (\ref{metric_eom}) 
and integrating it across the brane,
we obtain the junction condition for the metric
\bea\
\label{metric}
&&\Big[
-\Big\{1
+\frac{1}{2}
z\kappa^{\frac{2D}{D-2}}
\Big(\phi'{}^2-\phi^{|\rho}\phi_{|\rho}\Big)\Big\}
\big(K_{\mu\nu}- q_{\mu\nu} K\big)
\nonumber\\
&+&z \kappa^{\frac{2D}{D-2}}
\Big\{
-\phi_{|\mu\nu} 
\phi'
+\phi_{|\mu}\phi_{|\nu}K
-\phi^{|\alpha} \phi_{|\mu} K_{\alpha\nu}
-\phi^{|\alpha} \phi_{|\nu} K_{\alpha\mu}
+
q_{\mu\nu}
\Big(
K_{\rho\sigma} \phi^{|\rho} \phi^{|\sigma}
+\phi'
{}^{(q)}\Box \phi
\Big)
\Big\}
\Big]^+_-
=\kappa^2S_{\mu\nu}.
\eea
In the Einstein gravity limit $z=0$, we recover the Israel junction condition
\cite{israel}
\bea
\Big[
-K_{\mu\nu}+q_{\mu\nu} K\Big]^+_-
=\kappa^2S_{\mu\nu}.
\eea
Similarly, substituting \eqref{co3} into \eqref{scalar_eom}
and integrating it across the discontinuity, 
the scalar field junction condition is given by 
\bea
\label{scalar}
&&\Big[
\Big\{
1
+\frac{1}{2}
z\kappa^{\frac{4}{D-2}}
\Big(
{}^{(q)}R
+K^{\rho\sigma}K_{\rho\sigma}
-K^2
\Big)
\Big\}
\phi'
+
z\kappa^{\frac{4}{D-2}}
\big(
 K^{\mu\nu}
-q^{\mu\nu}K
\big)
\phi_{|\mu\nu}
\Big]^+_-
=-\frac{\delta {\cal L}_b}{\delta\phi}.
\eea
The equations of \eqref{cont}, \eqref{metric} and \eqref{scalar}
constitute a sef of the generalized junction conditions.
We notice that these junction conditions depend not only the extrinsic quantities $K_{\mu\nu}$ and $\phi'$
but also on the intrinsic quantities such as ${}^{(q)}R$ and $\phi_{|\mu}$.

Imposing the $Z_2$-symmetry across the brane $y=0$, 
namely $q_{\mu\nu}=q_{\mu\nu}(x^{\mu},|y|)$
and $\phi=\phi(x^{\mu},|y|)$,
the extrinsic curvature tensor and the scalar field derivative at $y=0+$ 
are given by 
\bea
\label{metric_z2}
&&
-\Big\{1
+\frac{1}{2}
z\kappa^{\frac{2D}{D-2}}
\Big(\phi'{}^2-\phi^{|\rho}\phi_{|\rho}\Big)
\Big\}
\big(K_{\mu\nu}- q_{\mu\nu} K\big)
\nonumber\\
&+&z\kappa^{\frac{2D}{D-2}}
\Big\{
-\phi_{|\mu\nu} 
\phi'
+\phi_{|\mu}\phi_{|\nu}K
-\phi^{|\alpha} \phi_{|\mu} K_{\alpha\nu}
-\phi^{|\alpha} \phi_{|\nu} K_{\alpha\mu}
+
q_{\mu\nu}
\Big(
K_{\alpha\beta} \phi^{|\alpha} \phi^{|\beta}
+\phi'
{}^{(q)}\Box \phi
\Big)
\Big\}
\Big|_{y=0+}
=\frac{\kappa^2}{2}S_{\mu\nu},
\eea
and 
\bea
\label{scalar_z2}
&&
\Big\{
1
+\frac{1}{2}
z\kappa^{\frac{4}{D-2}}
\Big(
{}^{(q)}R
+K^{\rho\sigma}K_{\rho\sigma}
-K^2
\Big)
\Big\}
\phi'
+z\kappa^{\frac{4}{D-2}}
\big(
 K^{\mu\nu}
-q^{\mu\nu}K
\big)
\phi_{|\mu\nu}
\Big|_{y=0+}
=-\frac{1}{2}\frac{\delta {\cal L}_b}{\delta\phi}.
\eea
In the particular case where the scalar field does not depend on the $(D-1)$-dimensional coordinate $x^\mu$,
namely
$\phi=\phi(y)$,
these boundary conditions reduce to 
\bea
\label{metric01}
&&
\Big(
 1
+
\frac{1}{2}
z\kappa^{\frac{2D}{D-2}}
(\phi')^2 
\Big)
\Big(
K_{\mu\nu}- q_{\mu\nu} K
\Big)
\Big|_{y=0+}
=-\frac{\kappa^2}{2}S_{\mu\nu},
\eea
and
\bea
\label{scalar01}
&&
\Big\{
1
+\frac{1}{2}
z\kappa^{\frac{4}{D-2}}
\Big(
{}^{(q)}R
+K^{\rho\sigma}K_{\rho\sigma}
-K^2
\Big)
\Big\}
\phi'
\Big|_{y=0+}
=-\frac{1}{2}\frac{\delta {\cal L}_b}{\delta\phi}.
\eea
Using \eqref{gyy}, \eqref{scalar01} can be rewritten as 
\bea
\label{gnn}
\Big(
1
-
z\kappa^{\frac{4}{D-2}}
G_{yy}
\Big)
\phi'
\Big|_{y=0+}
=-\frac{1}{2}\frac{\delta {\cal L}_b}{\delta\phi}.
\eea

The $(y,\mu)$ component of the gravitational equation of motion is given by 
\bea
G_{y\mu}+
z\kappa^{\frac{2D}{D-2}}
L_{y\mu}
=\kappa^2  \phi' \phi_{|\mu},
\label{eq}
\eea
where
\bea
G_{y\mu}&=&K^{\rho}{}_{\mu|\rho}-K_{|\mu},
\nonumber\\
L_{y\mu}
&=&
-\big(K\phi'+ {}^{(q)}\Box\phi \big)
 \big((\phi_{|\mu})' -K_{\mu}{}^{\rho}\phi_{|\rho}\big)
+q^{\alpha\beta}
\big(\phi_{|\alpha\mu} +K_{\alpha\mu} \phi'\big)
  \big((\phi_{|\beta})' -K_{\beta}{}^{\rho}\phi_{|\rho}\big)
-\frac{1}{2} \phi_{|\mu} \phi' \big(K^{\rho\sigma}K_{\rho\sigma}-K^2\big)
\nonumber\\
&+&\big(K_{\alpha\mu|\beta}-K_{\alpha\beta|\mu}\big) \phi^{|\alpha}\phi^{|\beta}
+\big(K^{\rho}{}_{\alpha|\rho} -K_{|\alpha}\big) \phi_{|\mu}\phi^{|\alpha}
-\frac{1}{2}\big(K^\rho{}_{\mu|\rho}-K_{|\mu}\big)
\phi^{|\alpha}\phi_{|\alpha}
\nonumber\\
&+&\frac{1}{2}\big(K^{\rho}{}_{\mu|\rho} -K_{|\mu}\big)(\phi')^2
+\phi' \phi^{|\nu}
\Big(
 {}^{(q)}G_{\mu\nu}
+K_{\mu\rho}K^{\rho}{}_{\nu}
-KK_{\mu\nu}
\Big).
\label{compymu}
\eea
In the case where the scalar field does not depend on the brane coordinate, $\phi_{|\mu}=0$,
from (\ref{metric01}) we obtain
\bea
0
&=&
\Big(1+\frac{1}{2}
z\kappa^{\frac{2D}{D-2}}
(\phi')^2\Big)\big(K^{\rho}{}_{\mu|\rho} -K_{|\mu}\big)
=-\frac{\kappa^2}{2}
 S_{\mu\nu}{}^{|\nu},
\eea
and hence 
the energy-momentum tensor of the matter on the brane is conserved, $S_{\mu\nu}{}^{|\nu}=0$.
On the other hand, if the scalar field also depends on the brane coordinate $x^{\mu}$,
there may be the energy exchange between the matter localized on the brane
and the bulk scalar field.

\section{The Minkowski and de Sitter brane solutions}

In the rest of the paper, we will focus on the case of $D=5$
and apply the formula obtained in the previous sections to 
the construction of the braneworld solutions in the $Z_2$ symmetric AdS bulk.
In this section, we derive the exact braneworld solutions embedded in the five-dimensional AdS
spactime. We assume a negative bulk cosmological constant $\Lambda<0$
(a constant negative bulk potential $V(\phi)=\frac{\Lambda}{\kappa^2}$
in the equations of motion \eqref{metric_eom} and \eqref{scalar_eom}),
and no matter field other than the scalar field in the bulk ${\cal L}_M=0$.
In general, it is difficult to obtain the exact solutions of the equations of motion 
in the bulk \eqref{metric_eom} and \eqref{scalar_eom} for a nonconstant bulk potential $V(\phi)$.
Because of our choice of the constant potential, despite the presence of the bulk scalar field, 
its contribution appears as the effective bulk cosmological constant.
To obtain the braneworld solutions for a noncontant bulk potential $V(\phi)$ will be left for the future work.

\subsection{The AdS spacetime}

In this section,
we assume the following ansatz for the bulk metric and the scalar field
\bea
\label{gingin}
ds^2=dy^2 +a(y)^2 \gamma_{\mu\nu}dx^{\mu}dx^{\mu},
\quad 
\phi=\phi(y),
\eea
where $a(y)$ represents the warp factor and $\gamma_{\mu\nu}$ does the maximally symmetric four-dimensional spacetime
with the non-negative constant curvature $R_{\mu\nu}[\gamma]=3H^2\gamma_{\mu\nu}$.
If $\gamma_{\mu\nu}$ represents the four-dimensional de Sitter spacetime,
the positive $H$ represents the constant Hubble expansion rate of the de Sitter spacetime.
The nonzero components of the curvature tensors associated with the metric \eqref{gingin}
are listed in Appendix A.
Using Eqs. \eqref{k1}, \eqref{k2} and \eqref{k3},
the $(y,y)$ and the trace of $(\mu,\nu)$ components of the gravitational equations of motion \eqref{metric_eom}
are explicitly given by 
\bea
\label{grav_maxima}
&&
6
\Big[\Big(\frac{a'}{a}\Big)^2
-\frac{H^2}{a^2}
\Big]
-\frac{1}{2}\kappa^2 (\phi')^2+\Lambda
+3 z\kappa^{\frac{10}{3}}
\Big[
3\Big(\frac{a'}{a}\Big)^2
-\frac{H^2}{a^2}
\Big]
(\phi')^2
=0,\\
&&
\label{grav_maxima2}
3\Big[\Big(\frac{a'}{a}\Big)^2
+\frac{a''}{a}
-\frac{H^2}{a^2}
\Big]
+\frac{1}{2}\kappa^2(\phi')^2
+\Lambda
+\frac{3}{2}z\kappa^{\frac{10}{3}}
\Big\{
\Big[
\Big(\frac{a'}{a}\Big)^2
+\frac{a''}{a}
+\frac{H^2}{a^2}
\Big](\phi')^2
+\frac{a'}{a}
\big[\big(\phi'\big)^2\big]'
\Big\}=0.
\eea
All the other components of the gravitational equations are trivially satisfied.
From Eq. (\ref{k4}),
the scalar field equation of motion \eqref{scalar_eom}
(multiplied by $\phi'$) is given by 
\bea
\label{sca_maxima}
\Big\{
1
-6z\kappa^{\frac{4}{3}}
\Big[
\Big(\frac{a'}{a}\Big)^2
-\frac{H^2}{a^2}
\Big]
\Big\}
\big[(\phi')^2\big]'
+\frac{8a'}{a}
\Big\{
1
-3z\kappa^{\frac{4}{3}}
\Big[
\frac{a''}{a}
+\Big(\frac{a'}{a}\Big)^2
-\frac{H^2}{a^2}
\Big]
\Big\}
(\phi')^2
=0.
\eea
After solving the equations \eqref{grav_maxima2} and \eqref{sca_maxima},
we will check the consistency with the constraint \eqref{grav_maxima}.
In the next subsubsections,
we will derive the AdS solution with the sections of the four-dimensional Minkowski and de Sitter spacetime.

\subsubsection{The AdS solution with the section of the four-dimensional Minkowski spacetime}

We derive the solutions of the AdS spacetime where $\gamma_{\mu\nu}$ in \eqref{gingin}
represents the four-dimensional Minkowski spacetime. 
We assume the vanishing Hubble expansion rate $H=0$ and the form of the warp factor
\bea
\label{rikudou}
a(y)=e^{-\frac{y}{\ell}},
\eea
where $\ell$ represents the curvature scale of the five-dimensional AdS spacetime. 
Substituting Eq. (\ref{rikudou})
into the scalar field equation of motion \eqref{sca_maxima}, we obtain 
\bea
\label{king}
\big(\ell^2-6z\kappa^{\frac{4}{3}}\big)
\Big[
-8(\phi')^2
+\ell\big((\phi')^2\big)'
\Big]=0.
\eea
There are two cases which satisfy the scalar field equation of motion \eqref{king},
(a): 
\bea
\label{ell5}
\ell=\sqrt{6z}\kappa^{\frac{2}{3}}.
\eea 
and (b): $-8(\phi')^2 +\ell\big((\phi')^2\big)'=0$.

First, we consider the case (a).
Eq. (\ref{ell5}) gives $z>0$ and then \eqref{grav_maxima2} reduces to
\bea
\frac{1}{z\kappa^{\frac{4}{3}}}
+\Lambda
+\kappa^2 (\phi'(y))^2
-\frac{1}{2}\sqrt{\frac{3z}{2}} \kappa^{\frac{8}{3}}
 \big[(\phi'(y))^2\big]'
=0,
\eea
which can be integrated with respect to $y$ as 
\bea
(\phi'(y))^2
=-\frac{6}{\kappa^2\ell^2}\big(1+\Lambda z\kappa ^{\frac{4}{3}}\big)
+ D_1 e^{\frac{4y}{\ell}},
\eea
where $D_1$ is an integration constant.
From the remaining \eqref{grav_maxima}, we find the condition 
$D_1 e^{\frac{4y}{\ell}}=0$, which allows the choice $D_1=0$. 
Since $(\phi')^2$ has to be non-negative,
we have to impose 
\bea
\label{condle}
1+\Lambda z \kappa^{\frac{4}{3}}
 \leq 0.
\eea
Integrating $\phi'(y)$ with respect to $y$ once more, with the boundary condition $\phi(0)=0$,
we obtain the AdS solution with the section of the Minkowski spacetime
\bea
\label{min5}
a(y)= e^{-\frac{y}{\ell}}, \quad 
\phi(y)=\pm \sqrt{-6(1+\Lambda z\kappa^{\frac{4}{3}})} \frac{y}{\kappa\ell},
\quad \ell=\sqrt{6z}\kappa^{\frac{2}{3}}.
\eea
Note that the solution Eq. (\ref{min5}) does not require the tuning relation between $\Lambda$ and $z$,
except for the conditions $z>0$ and \eqref{condle}.
Also,  without loss of the generality, in the discussions below, we will choose the negative branch of $\phi(y)$.

Next, we consider the case (b),
which gives $(\phi(y)')^2= D_2 e^{\frac{8y}{\ell}}$ after the integration with respect to $y$,
 where $D_2$ is an integration constant.
The gravitational equations of motion \eqref{grav_maxima} and \eqref{grav_maxima2} then reduce to
\bea
&&D_2\,
e^{\frac{8y}{\ell}}\kappa^2
\big(
-\ell^2+18 z\kappa^{\frac{4}{3}}
\big)
+2 (6+\Lambda \ell^2)=0,
\quad 
D_2\,
e^{\frac{8y}{\ell}}\kappa^2
\big(
\ell^2-18 z\kappa^{\frac{4}{3}}
\big)
+2 (6+\Lambda \ell^2)=0.
\eea
The only possibility to satisfy both of them for an arbitrary $y$
is to impose
\bea
\label{tuning_rel}
\ell=3\sqrt{2z}\kappa^{\frac{2}{3}},
\quad
\Lambda=-\frac{1}{3z\kappa^{\frac{4}{3}}}
=-\frac{6}{\ell^2}.
\eea
Integrating $\phi'(y)$ once more with the boundary condition $\phi(0)=0$,
\bea
\label{tuning_sol}
\phi(y)=\pm \frac{\ell}{4}\sqrt{D_2} \Big(e^{\frac{4y}{\ell}}-1\Big).
\eea
The solution \eqref{tuning_sol} 
requires the tuning relation between $\Lambda$ and $z$ \eqref{tuning_rel}.
The value $\Lambda z\kappa^{\frac{4}{3}}=-\frac{1}{3}$ is out of 
the condition imposed in \eqref{condle}.
In the discussions below, we will focus on case that the condition \eqref{condle} is satisfied
and will not consider the solution of \eqref{tuning_sol}.


\subsubsection{The AdS solution with the section of the four-dimensional de Sitter spacetime}

We then derive the solutions of the AdS spacetime where $\gamma_{\mu\nu}$ in \eqref{gingin}
represents the four-dimensional de Sitter spacetime. 
We assume the positive Hubble expansion rate $H>0$ and the form of the warp factor
\bea
\label{rikudou2}
a(y)=H\ell\sinh \Big(\frac{y_0-y}{\ell}\Big),
\eea
where $\ell$ represents the curvature scale of the AdS spacetime and $y_0$ is a constant.
Here we assume that $0<y<y_0$ and the position $y=y_0$ corresponds to the AdS horizon. 
Substituting Eq. (\ref{rikudou2}) into the scalar field equation of motion \eqref{sca_maxima},
we obtain 
\bea
\big(\ell^2-6z\kappa^{\frac{4}{3}}\big)
\Big[
-8\coth\big(\frac{y_0-y}{\ell}\big) (\phi')^2
+\ell\big((\phi')^2\big)'
\Big]=0. \label{tensei}
\eea
To satisfy Eq. (\ref{tensei}), again there are two cases,
(a): Eq. (\ref{ell5})
and
(b): $-8\coth\big(\frac{y_0-y}{\ell}\big) (\phi')^2
+\ell\big((\phi')^2\big)'=0$.
The case (a) is very similar to that in the previous subsubsection.

First, we consider the case (a).
Eq. (\ref{ell5}) imposes $z>0$ and 
then the gravitational equation of motion \eqref{grav_maxima2} reduces to
\bea
\frac{1}{z\kappa^{\frac{4}{3}}}
+\Lambda
+\frac{\kappa^2}{2}\cosh\Big(\frac{2}{\ell} (y_0-y)\Big)
\sinh^{-2}\Big(\frac{1}{\ell} (y_0-y)\Big)(\phi'(y))^2
-\frac{1}{2}\sqrt{\frac{3z}{2}} \kappa^{\frac{8}{3}}
\coth \Big(\frac{1}{\ell} (y_0-y)\Big)
 \big\{ (\phi'(y))^2\big\}'
=0,
\eea
which can be integrated with respect to $y$ as 
\bea
(\phi'(y))^2
=-\frac{6}{\kappa^2\ell^2}\big(1+\Lambda z\kappa ^{\frac{4}{3}}\big)
 \tanh^2 \Big(\frac{1}{\ell} (y_0-y)\Big)
+ \tilde D_1 \sinh^{-2}\Big(\frac{2}{\ell} (y_0-y)\Big),
\eea
where $\tilde D_1$ is an integration constant.
From the remaining \eqref{grav_maxima}, we find 
${\tilde D}_1  \sinh^{-4}\big(\frac{1}{\ell} (y_0-y)\big)=0$, which gives $\tilde D_1=0$. 
Since $(\phi')^2$ has to be non-negative, we have to impose \eqref{condle}.
Integrating $\phi'(y)$ once more by setting the boundary condition $\phi(0)=0$,
we obtain the AdS solution 
\bea
\label{ds5}
&&a(y)= H\ell\sinh\Big(\frac{y_0-y}{\ell}\Big), \quad 
\phi(y)=\pm \sqrt{-6(1+\Lambda z\kappa^{\frac{4}{3}})} 
\frac{1}{\kappa}
\ln\Big(\frac{\cosh \big(\frac{y_0-y}{\ell}\big)}{\cosh \big(\frac{y_0}{\ell}\big)}\Big),
\quad
\ell=\sqrt{6z}\kappa^{\frac{2}{3}}.
\eea
Again, the solution Eq. (\ref{ds5}) does not require the tuning relation between $\Lambda$ and $z$,
except for the conditions $z>0$ and \eqref{condle}.
Also,  without loss of the generality, in the discussions below, we will choose the negative branch of $\phi(y)$.


Next, we consider the case (b), which gives $(\phi(y)')^2= \tilde D_2 \sinh^{-8}\big(\frac{1}{\ell} (y_0-y) \big)$ after the integration with respect to $y$,
where $\tilde D_2$ is an integration constant.
The gravitational equations of motion
\eqref{grav_maxima} and \eqref{grav_maxima2} then reduce to 
\bea
&&2(6+\ell^2\Lambda)
+\kappa^2\tilde D_2
\sinh^{-8}
\Big(\frac{1}{\ell} (y_0-y)\Big)
\Big[
-\ell^2+18 z\kappa^{\frac{4}{3}}
+12 z\kappa^{\frac{4}{3}}
\sinh^{-2}
\Big(\frac{1}{\ell} (y_0-y)\Big)
\Big]
=0,
\nonumber\\
&&2(6+\ell^2\Lambda)
+\kappa^2\tilde D_2
\sinh^{-8}
\Big(\frac{1}{\ell} (y_0-y)\Big)
\Big[
\ell^2-18 z\kappa^{\frac{4}{3}}
-18 z\kappa^{\frac{4}{3}}
\sinh^{-2}
\Big(\frac{1}{\ell} (y_0-y)\Big)
\Big]
=0,
\eea
which are incompatible for an arbitrary $y$.
Thus there is no AdS solution from the case (b),
in contrast to the case of the section of the Minkowski spacetime
discussed in the subsubsection IV-A-1.


\vspace{.2cm}

Therefore, 
the solutions Eqs. \eqref{min5} and \eqref{ds5} give the general AdS solutions
for the sections of the four-dimensional Minkowski and de Sitter spacetime, respectively,
in the sense that they do not require the tuning relation between $z$ and $\Lambda$ (except for the restrictions (32) and $z>0$).
Hence, in the rest of this section,
we will focus on the AdS solutions \eqref{min5} and \eqref{ds5}
for obtaining the Minkowski and de Sitter brane solutions, respectively.
In these solutions \eqref{min5} and \eqref{ds5} obtained from Eq. (\ref{ell5}),
the five-dimensional AdS spacetime satisfies 
\bea
G_{AB}=\frac{1}{z\kappa^{\frac{4}{3}}}g_{AB}
=-\Lambda_{\rm eff}g_{AB}.
\label{gg}
\eea
The effective cosmological constant becomes negative $\Lambda_{\rm eff}<0$,
because of $z>0$.

In our solutions the effect of the bulk scalar field with the field derivative coupling to the Einstein tensor appears as the effective cosmological constant $\Lambda_{\rm eff}$.
This is the case that because of the constant potential $V(\phi)=\frac{\Lambda}{\kappa^2}$, $\frac{dV}{d\phi}=0$
and the scalar field equation motion \eqref{scalar_eom} is trivially satisfied. 
For a nonconstant potential $V(\phi)$, $\frac{dV}{d\phi}\neq 0$ and solving the scalar field equation of motion \eqref{scalar_eom} would be a more involved problem.

The braneworld solutions can be realized in the $Z_2$-symmetric AdS background, 
obtained by replacing $y$ with $|y|$ in 
the pure AdS solutions Eqs. \eqref{min5} and \eqref{ds5}.

\subsubsection{The junction conditions}


Next, we focus on the junction conditions.
From \eqref{gnn} and \eqref{gg}, we find $\frac{\delta {\cal L}_b}{\delta\phi}=0$.
Thus for our construction,
the scalar field is not coupled to the brane,
and the maximally symmetric brane can be supported by the constant tension
\bea
\label{blp}
{\cal L}_b= -\sigma,  
\eea
where $\sigma={\rm const.}$ is the brane tension.
Therefore, we just need to solve \eqref{metric01} to determine the brane tension.

\subsection{The Minkowski brane solutions}

We first consider the Minkowski braneworld solution, 
given by $\gamma_{\mu\nu}=\eta_{\mu\nu}$ and the warp factor in \eqref{min5} with $y\to |y|$,
\bea
\label{min_ads}
a(y)= e^{-\frac{|y|}{\ell}},
\eea
The absolute value for the bulk coordinate $|y|$ reflects the $Z_2$-symmetry across the brane at $y=0$.
Without loss of generality, 
the configuration of the scalar field is given
by the negative branch of Eq. (\ref{min5}) with $y\to |y|$,
\bea
\label{config}
\phi(y)=-\sqrt{-6(1+\Lambda z \kappa^{\frac{4}{3}} )} \frac{|y|}{\kappa \ell},
\eea
where the integration constant is chosen to be $\phi(0)=0$.
In order to make configuration of the scalar field to be physical,
we have to impose the condition 
\eqref{condle}.
which leads to 
$\Lambda\leq \Lambda_{\rm eff}$ or equivalently $\Lambda\leq-\frac{6}{\ell^2}$.
We notice that the equality in \eqref{condle} holds for the case where the scalar field becomes trivial.

The metric junction \eqref{metric01} gives the brane tension
\bea
\label{tension_min}
\kappa^2\sigma=\frac{6}{\ell} 
\Big[
1-\frac{1}{2}\big(
1+\Lambda z \kappa^{\frac{4}{3}}
\big)
\Big].
\eea
From \eqref{condle}, we find that $\sigma>0$.

\subsection{The de Sitter brane solutions}

Next, we focus on the de Sitter braneworld solutions with 
the expansion rate $H$, where the warp factor is given by Eq. (\ref{ds5})
with $y\to |y|$,
\bea
a(y)&=&H\ell\sinh \Big(\frac{y_0-|y|}{\ell}\Big).
\eea
The bulk region of the spacetime is $-y_0<y<y_0$
and $y=\pm y_0$ corresponds to the bulk AdS horizon.
The scalar field equation of motion is automatically satisfied.

Without loss of generality, 
the configuration of the scalar field is given by 
by the negative branch of Eq. (\ref{ds5}) with $y\to |y|$,
\bea
\phi(y)&=&
\sqrt{-6(1+\Lambda z \kappa^{\frac{4}{3}})
}
\frac{1}{\kappa}
\ln \Big(\frac{\cosh \big(\frac{y_0-|y|}{\ell}\big)}{\cosh \big(\frac{y_0}{\ell}\big)}\Big),
\eea
where the integration constant is chosen to be $\phi(0)=0$.
We require that 
at the brane position $y=0$ so that
$\gamma_{\mu\nu}$ represents the physical four-dimensional metric on the brane
$H\ell \sinh\Big(\frac{y_0}{\ell}\Big)=1$.
As for the Minkowski brane solution,
we have to impose the condition Eq. (\ref{condle}).
The scalar field becomes trivial for $1+\Lambda z \kappa^{\frac{4}{3}}=0$.

The metric boundary condition \eqref{metric01} gives the brane tension
\bea
\label{hagen}
\kappa^2\sigma=\frac{6}{\ell} 
\sqrt{1+(H\ell)^2}
\Big[
1-\frac{1}{2}\frac{1+\Lambda z \kappa^{\frac{4}{3}}
}{1+(H\ell)^2}
\Big].
\eea
From \eqref{condle}, we always have $\sigma>0$. 
Solving Eq. (\ref{hagen}) in terms of $H$,
\bea
\label{pm}
\sqrt{1+ H_\pm^2\ell^2}
=\frac{1}{2}
\Big(
\frac{\kappa^2\sigma\ell}{6}
\pm
\sqrt{\frac{\kappa^4\sigma^2\ell^2}{36}+2(
1+\Lambda z \kappa^{\frac{4}{3}}
)}
\Big),
\eea
where $H_{+}$ and $H_-$ take the upper and lower branches, respectively. 
In order for $H_{\pm}^2$ to be real,
the quantity inside the square root in the right-hand side of \eqref{pm}
must be positive, leading to the lower bound on the brane tension,
\bea
\label{bound}
\kappa^2\sigma
\geq
\frac{6}{\ell}
\sqrt{-2\big(
1+\Lambda z \kappa^{\frac{4}{3}}
\big)}.
\eea
From \eqref{pm}, it is clear that as the brane tension increases,
$H_+$ increases while $H_-$ decreases.
The condition for obtaining the physical solution
is that the right-hand side of \eqref{pm} must be greater than unity.
To make things more explicit,
let us investigate the behavior of the function
\bea
F_{\pm}(X)=\frac{1}{2}
\Big(
X
\pm
\sqrt{X^2-Y^2}
\Big),
\eea
where we have defined the dimensionless quantities
\bea
X:=\frac{\kappa^2\sigma\ell}{6},\quad
Y:=\sqrt{-2(1+\Lambda z \kappa^{\frac{4}{3}}
)}.
\eea
We notice that the functions $F_{\pm}(X)$ are defined for $X\geq Y$, 
which is equivalent to \eqref{bound}.
The behaviors of $F_{\pm}(X)$ are shown in Fig. 1.
\begin{figure}[tbph]
\begin{center}
\begin{tabular}{ll}
\includegraphics[width=.5\linewidth,origin=tl]{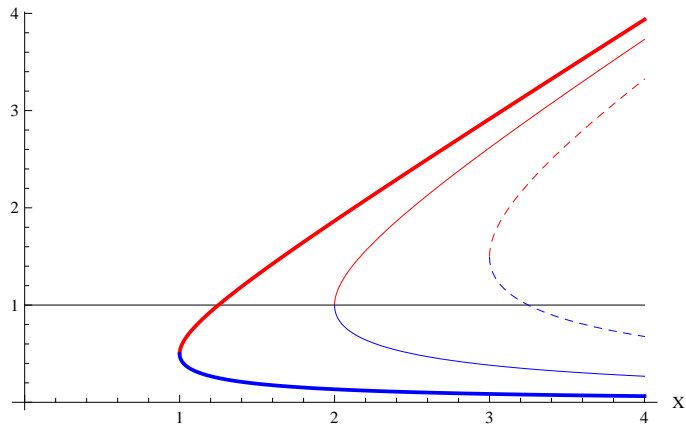}
\end{tabular}
\end{center}
\caption{
The functions $F_\pm(X)$ are shown as the function of $X$.
The red and blue curves correspond to $F_+$ and $F_-$, respectively.
For each case, the thick, solid and dashed curves correspond to $Y=1,2,3$, respectively.
The black horizontal line shows the threshold value $F_{\pm}=1$.
} \label{fig:p4}
\end{figure}
For $Y<2$, $F_+(X)$ is greater than unity for $X>1+\frac{1}{4}Y^2$,
while for $Y>2$ $F_+(X)$ is always greater than unity.
On the other hand, if $Y>2$,
$F_-(X)$ can be greater than unity for $Y<X<1+\frac{1}{4}Y^2$.
From these properties, we can discuss the existence of the solutions.

\subsubsection{The $H_+$ solution}

The $H_+$ solution can always provide the physical de Sitter spacetime
for $
1+\Lambda z \kappa^{\frac{4}{3}}
<-2$.
On the other hand, in the opposite case that
$1+\Lambda z \kappa^{\frac{4}{3}}>-2$,
the solution can provide the de Sitter spacetime 
if the brane tension satisfies 
\bea
\kappa^2\sigma >\frac{6}{\ell}\Big[1-\frac{1}{2}(
1+\Lambda z \kappa^{\frac{4}{3}}
)\Big].
\eea
The latter case also covers the condition for the de Sitter brane solutions in the case with a trivial scalar field,
$\kappa^2\sigma >\frac{6}{\ell}$.

\subsubsection{The $H_-$ solution}

In order to make the $H_-$ solution to be physical,
we require $1+\Lambda z \kappa^{\frac{4}{3}}<-2$
and
\bea
\label{bound2}
\kappa^2\sigma<\frac{6}{\ell}\Big[1-\frac{1}{2}\big(
1+\Lambda z \kappa^{\frac{4}{3}}
\big)\Big].
\eea
Thus the $H_-$ solution does not contain the smooth limit 
to the case with a trivial scalar field $1+\Lambda z \kappa^{\frac{4}{3}}\to 0$.
This branch has the smaller expansion rate $H_-<H_+$
for a given brane tension,
and for a larger brane tension, $H_-$ decreases.
When the upper bound \eqref{bound2} is saturated, $H_-$ vanishes.


\section{The low energy effective theory on the brane}

Before closing this paper, using the gradient expansion approach \cite{ge,ge1,ge2},
we derive the effective gravitational theory realized on the brane at the energy low regime.
Here we consider the situation 
that the brane intrinsic curvature is much smaller than that of the bulk AdS spacetime.
For simplicity, again, we assume the $Z_2$-symmetry across the brane, ${\cal L}_M=0$
and $V(\phi)=\frac{\Lambda}{\kappa^2}$.

\subsection{The expansion scheme}

We define the small parameter which characterizes the low energy expansion of the bulk degrees of freedom
\bea
\epsilon=\big|{}^{(h)}R\big| \ell^2(\ll1),
\eea
where ${}^{(h)}R$ is the intrinsic scalar curvature on the brane 
and $\ell$ is the bulk AdS curvature scale given below.
We assume the relation between the extrinsic and intrinsic derivatives 
\bea
\partial_{\mu}A \sim \epsilon^{\frac{1}{2}} \partial_y A.
\eea
We expand the bulk metric in terms of $\epsilon$ as
\bea
q_{\mu\nu}(x^{\mu},y)&=& a^2(y)\Big[ h_{\mu\nu}(x^{\mu})+q^{(1)}_{\mu\nu}(x^\mu,y)+q^{(2)}_{\mu\nu}(x^\mu,y)+\cdots\Big],
\eea
where $A^{(n)}$ is the quantity of $O(\epsilon^n)$.
The extrinsic curvature tensor is also expanded as
\bea
K^{\mu}{}_{\nu}
&=&K^{(0)\mu}{}_{\nu}
+K^{(1)\mu}{}_{\nu}
+K^{(2)\mu}{}_{\nu}
+\cdots.
\eea
Without the loss of generality, we can set boundary conditions on the brane
$a(0)=1$ and $q^{(i)}_{\mu\nu}(x^\mu,0)=0$ ($i=1,2,3,\cdots$), 
hence the physical metric on the brane is given by $q_{\mu\nu}(x^{\mu},0)=h_{\mu\nu}(x^{\mu})$.
In this section,
we decompose the extrinsic curvature into the trace and traceless components,
\bea
K^{\mu}{}_{\nu}=\frac{1}{4}K \delta^{\mu}{}_{\nu}+\Sigma^{\mu}{}_{\nu},
\quad 
K=
\frac{1}{\sqrt{-q}}
\frac{\partial}{\partial y}
(\sqrt{-q}),
\quad
\Sigma^{\mu}{}_{\mu}=0.
\eea
Similarly, the scalar field is expanded as
\bea
\phi&=& \phi^{(0)} +\phi^{(1)}+\phi^{(2)}+\cdots,
\eea
where we require the boundary condition $\phi(x^{\mu},0)=\varphi(x^{\mu})$
so that $\varphi(x^{\mu})$ becomes the boundary value of the scalar field.
In this way, expanding the equations of motion \eqref{metric_eom} and \eqref{scalar_eom} (except for the $\delta(y)$ terms)
in terms of $\epsilon$,
at each order
we solve them with the given boundary conditions on the brane 
obtained from the corresponding expansion of Eqs. \eqref{metric_z2} and \eqref{scalar_z2}.
In this paper, for simplicity,
we will focus on the solutions at the zeroth and first orders.
In our case
it is interesting to see how the contribution of the bulk scalar field
gives the deviation from the four-dimensional Einstein gravity.

The energy-momentum tensor on the brane is given by
\bea
S^{\mu}{}_{\nu}&=&-\sigma\delta^{\mu}{}_\nu + \tau^{\mu}{}_{\nu}(x^{\mu}),
\eea
where $\sigma$ is the brane tension
and $\tau{}^{\mu}{}_{\nu}$ is the energy-momentum tensor
of the ordinary matter localized on the brane.
As we will see,
the zeroth order part of the junction condition determines the tension
and the the first order part determines $\tau^{\mu}{}_{\nu}$. 
The variation of the brane Lagrangian density with respect to the scalar field is denoted by
\bea
\label{kim}
U(\varphi)
:=\frac{\delta {\cal L}_b}{\delta\phi}
\Big|_{\phi=\varphi}.
\eea

\subsection{The zeroth order solution}

The zeroth order part of the traceless part of the $(\mu,\nu)$ component of the bulk
gravitational equation of motion \eqref{metric_eom}
is given by
\bea
\label{zero_traceless}
\Big(1+\frac{1} {2}z\kappa^{\frac{10}{3}}(\phi^{(0)}{}')^2 \Big)
\Big(
\Sigma^{(0)}{}^{\mu}{}_{\nu,y}
+K^{(0)}\Sigma^{(0)}{}^{\mu}{}_{\nu}
\Big)
+\frac{1}{2}z\kappa^{\frac{10}{3}} \big((\phi^{(0)}{}')^2\big)'\Sigma^{(0)}{}^{\mu}{}_{\nu}
&=&0.
\eea
Imposing $(\phi^{(0)}{})''=0$ and integrating \eqref{zero_traceless},
we obtain
\bea
\Sigma^{(0)}{}^{\mu}{}_{\nu}=\frac{F^{\mu}{}_{\nu}(x^{\mu})} {\sqrt{-q}},
\eea
where $F^{\mu}{}_{\nu}(x^{\mu})$ is the traceless integration constant.
We require that there is no anisotropy on the brane at the level of the zeroth order
and hence set $F^{\mu}{}_{\nu}(x^{\mu})=0$.
The zeroth order part of the remaining components of the bulk
gravitational equation of motion \eqref{metric_eom}
is then given by
\bea
\label{zero}
\Big(\frac{1}{2}+\frac{3} {4}z\kappa^{\frac{10}{3}}(\phi^{(0)}{}')^2 \Big)
\frac{3}{4}\big(K^{(0)}\big){}^2
&=&\frac{1}{2}\kappa^2(\phi^{(0)}{}')^2-\Lambda,
\nonumber\\
\Big(1+\frac{1} {2}z\kappa^{\frac{10}{3}}(\phi^{(0)}{}')^2 \Big)
\Big(
3K^{(0)}_{,y}
+\frac{3}{2}\big(K^{(0)}\big){}^2
\Big)
&=&-4\Big\{\frac{1}{2}\kappa^2(\phi^{(0)}{}')^2+\Lambda\Big\}.
\eea
Assuming $K^{(0)}_{,y}=0$
they can be algebraically solved as 
\bea
\label{zeroth_1st}
K^{(0)}=-\frac{4}{\ell},\quad 
\phi^{(0)}{}'
=-\sqrt{-6\Big(1+\Lambda z\kappa^{\frac{4}{3}}\Big)}\frac{1}{\kappa\ell}.
\eea
where $\ell=\sqrt{6z}\kappa^{\frac{2}{3}}$ is being the AdS curvature scale.
From the definition of the extrinsic curvature,
\bea
K^{(0)}_{\mu\nu}=\frac{1}{2} \Big(\frac{\partial}{\partial y} a(y)^2\Big) h_{\mu\nu}(x^{\mu}),
\eea
and integrating \eqref{zeroth_1st} with respect to $y$ once more and setting the integration constants 
to satisfy the boundary conditions on the brane $a(0)=1$ and $\phi^{(0)}=\varphi(x^{\mu})$,
the bulk metric and the configuration of the scalar field at the zeroth order 
are found to be 
\bea
\label{0thsol}
ds^2&=& dy^2+ a(y)^2 h_{\mu\nu}(x^{\mu})dx^{\mu}dx^{\nu},
\nonumber\\
a(y)&=& e^{-\frac{|y|}{\ell}},
\nonumber\\
\phi^{(0)}&=&-\sqrt{-6(1+\Lambda z \kappa^{\frac{4}{3}} )} \frac{|y|}{\kappa \ell}+\varphi(x^{\mu}),
\eea
where the absolute value $|y|$ reflects the $Z_2$ symmetry across the brane.
The bulk metric also gives the spacetime curvature at the zeroth order
\bea
G^{(0)}
{}^{A} {}_{B}
=\frac{1}{z\kappa^{\frac{4}{3}}}
\delta  {}^{A} {}_{B},
\label{gg2}
\eea
for which the zeroth order part of the bulk scalar field equation of motion \eqref{scalar_eom}  is automatically satisfied.
We have to impose the same condition as Eq. (\ref{condle}).
It is also straightforward to check 
that the $O(\epsilon^{\frac{1}{2}})$ part of the $(y,\mu)$ component of the gravitational equation \eqref{eq}
is satisfied.

We then investigate the junction conditions on the brane.
The zeroth order part of the metric boundary condition \eqref{metric_z2} gives the same brane tension as \eqref{tension_min}.
The zeroth order part of the scalar field boundary condition \eqref{scalar_z2} 
gives $U(\varphi)=O(\epsilon)$,
because the left-hand side of \eqref{scalar_z2} vanishes at the zeroth order. 
Thus at the zeroth order level, 
the bulk scalar field is not coupled to the brane.

\subsection{The effective gravitational theory on the brane}

In this subsection, we investigate the solution to the bulk equations of motion
at the first order of the gradient expansion
and derive the effective gravitational equations on the brane at the low energy regimes.
At the first order equations of motion,
we may approximate
${}^{(q)}R^{\mu}{}_{\nu}=\frac{1}{a(y)^2}{}^{(h)}R^{\mu}{}_{\nu}(x^{\mu})$ in the bulk,
where ${}^{(h)}R^{\mu}{}_{\nu}$ is the Ricci tensor for the brane metric $h_{\mu\nu}(x^{\mu})$.
From the discussions in the previous subsections,
we impose the boundary conditions on the brane 
$q^{(1)}_{\mu\nu}(x^{\mu},0)=0$ and 
$\phi^{(1)}(x^{\mu},0)=0$.

At the first order,
the traceless part of the $(\mu,\nu)$ component of the gravitational equation \eqref{metric_eom}
is given by
\bea
\label{traceless_eq}
&-&\frac{1}{a(y)^4}\Big\{a(y)^4
\Big(1-\frac{1}{2}\big(1+\Lambda\kappa^{\frac{4}{3}} z\big)\Big)
\Sigma^{(1)}{}^{\mu}{}_{\nu}
\Big\}_{,y}
+\frac{1}{a^2(y)}
\Big(1+\frac{1}{2}\big(1+\Lambda\kappa^{\frac{4}{3}} z\big)\Big)
\Big(
{}^{(h)}R^\mu{}_\nu(x^{\mu})
-\frac{1}{4}\delta^{\mu}{}_\nu{}^{(h)}R(x^{\mu})
\Big)
\nonumber\\
&+&\frac{\kappa^2\ell}{3a(y)^2}(\phi^{(0)})'
\Big(D^{\mu}D_{\nu}\varphi(x^{\mu})
-\frac{1}{4}\delta^{\mu}{}_{\nu}{}^{(h)}\Box \varphi(x^{\mu})
\Big)
\nonumber\\
&-&\frac{\kappa^2}{3a(y)^2}
\Big(D^{\mu} \varphi(x^{\mu}) D_{\nu}\varphi(x^{\mu})
-\frac{1}{4}\delta^{\mu}{}_{\nu}
D^\rho \varphi(x^{\mu})
D_\rho\varphi(x^{\mu})
\Big)
=0,
\eea
where $D_{\mu}$ is the covariant derivative with respect to $h_{\mu\nu}$
and ${}^{(h)}\Box:= h^{\mu\nu}D_{\mu}D_{\nu}$.
Integrating Eq. (\ref{traceless_eq}) with respect to $y$,
we obtain 
\bea
\label{beq1}
\Big(1-\frac{1}{2}\big(1+\Lambda\kappa^{\frac{4}{3}} z\big)\Big)
\Sigma^{(1)}{}^{\mu}{}_{\nu}
&=&-\frac{\ell}{2a^2(y)}
\Big(1+\frac{1}{2}\big(1+\Lambda\kappa^{\frac{4}{3}} z\big)\Big)
\Big(
{}^{(h)}R^\mu{}_\nu (x^{\mu})
-\frac{1}{4}\delta^{\mu}{}_\nu{}^{(h)}R (x^{\mu})
\Big)
\nonumber\\
&-&\frac{\kappa^2\ell^2 }{6a(y)^2}
 (\phi^{(0)})'
\Big(D^{\mu}D_{\nu} \varphi  (x^{\mu})
-\frac{1}{4}\delta^{\mu}{}_{\nu}{}^{(h)}\Box \varphi   (x^{\mu})\Big)
\nonumber\\
&+&\frac{\kappa^2\ell}{6a(y)^2}
\Big(D^{\mu}\varphi  (x^{\mu}) D_{\nu} \varphi (x^{\mu})
-\frac{1}{4}\delta^{\mu}{}_{\nu}D^{\rho} \varphi  (x^{\mu})
D_{\rho}\varphi  (x^{\mu})\Big)
+\frac{\ell}{2a(y)^4}\psi^{\mu}{}_\nu(x^{\mu}),
\eea
where $\psi^{\mu}{}_\nu(x^{\mu})$ is the integration constant satisfying
the traceless condition $\psi^{\mu}{}_{\mu}(x^{\mu})=0$.
Taking the limit to the brane $y\to 0$ with $a(0)=1$,
\bea
\label{traceless}
\Big(1-\frac{1}{2}\big(1+\Lambda\kappa^{\frac{4}{3}} z\big)\Big)
\Sigma^{(1)}{}^{\mu}{}_{\nu}
&=&-\frac{\ell}{2}
\Big(1+\frac{1}{2}\big(1+\Lambda\kappa^{\frac{4}{3}} z\big)\Big)
\Big(
{}^{(h)}R^\mu{}_\nu   (x^{\mu})
-\frac{1}{4}\delta^{\mu}{}_\nu{}^{(h)}R (x^{\mu})
\Big)
\nonumber\\
&-&\frac{\kappa^2\ell^2 }{6}
 (\phi^{(0)})'
\Big(D^{\mu}D_{\nu} \varphi (x^{\mu})
-\frac{1}{4}\delta^{\mu}{}_{\nu}{}^{(h)}\Box \varphi  (x^{\mu})
\Big)
\nonumber\\
&+&\frac{\kappa^2\ell}{6} 
\Big(D^{\mu}\varphi  (x^{\mu})D_{\nu} \varphi  (x^{\mu})
-\frac{1}{4}\delta^{\mu}{}_{\nu}D^{\rho} \varphi   (x^{\mu})
                                            D_{\rho}\varphi    (x^{\mu})
\Big)
+\frac{\ell}{2}\psi^{\mu}{}_\nu(x^{\mu}).
\eea
Similarly,
the first order part of the trace part of the $(\mu,\nu)$ component of the gravitational equation \eqref{metric_eom}
is given by 
\bea
\label{trace_eq}
&&\frac{3}{a(y)^4}
\Big\{
a(y)^4\Big[\Big(1-\frac{1}{2}\big(1+\Lambda\kappa^{\frac{4}{3}} z\big)\Big) K^{(1)}
-\frac{2}{3}\kappa^2\ell (\phi^{(0)})'(\phi^{(1)})'
\Big]
\Big\}_{,y}
\nonumber\\
&-&\frac{1}{a(y)^2}
\Big[
\Big(1+\frac{1}{2}\big(1+\Lambda\kappa^{\frac{4}{3}} z\big)\Big){}^{(h)}R (x^{\mu})
+\kappa^2\ell (\phi^{(0)})'
{}^{(h)}\Box \varphi (x^{\mu})
-\kappa^2 D^{\rho}\varphi    (x^{\mu})
              D_{\rho}\varphi     (x^{\mu})
\Big]=0,
\eea
which is integrated as 
\bea
\label{trace_sol}
&&\Big(1-\frac{1}{2}\big(1+\Lambda\kappa^{\frac{4}{3}} z\big)\Big) K^{(1)}
-\frac{2}{3}\kappa^2\ell (\phi^{(0)})' (\phi^{(1)})'
\nonumber \\
&=&-\frac{\ell}{6a(y)^2}
\Big[
\Big(1+\frac{1}{2}\big(1+\Lambda\kappa^{\frac{4}{3}} z\big)\Big){}^{(h)}R (x^{\mu})
+\kappa^2\ell (\phi^{(0)})'
{}^{(h)}\Box \varphi  (x^{\mu})
-\kappa^2 D^\rho \varphi  (x^{\mu}) D_\rho\varphi (x^{\mu})
\Big]
+\frac{C(x^{\mu})}{a(y)^4},
\eea
where $C(x^{\mu})$ is an integration constant. 
The $(y,y)$ component of the gravitational equation gives
the constraint condition in the bulk
\bea
\label{const_bulk}
\Big(1-\frac{3}{2}\big(1+\Lambda\kappa^{\frac{4}{3}} z\big)\Big)K^{(1)}
-\frac{2}{3}\kappa^2\ell (\phi^{(0)})'(\phi^{(1)})'
&=& 
-\frac{\ell}{6a(y)^2}
  \Big(1-\frac{1}{2}\big(1+\Lambda\kappa^{\frac{4}{3}} z\big)\Big){}^{(h)}R (x^{\mu})
-\frac{\kappa^2\ell^2}{6a(y)^2} (\phi^{(0)})' {}^{(h)}\Box\varphi (x^{\mu})
\nonumber\\
&+&\frac{\kappa^2\ell}{6a(y)^2}D^{\rho}\varphi  (x^{\mu})
                                            D_{\rho}\varphi (x^{\mu}),
\eea
whose projection onto the brane is given by
\bea
\label{1st2}
\Big(1-\frac{3}{2}\big(1+\Lambda\kappa^{\frac{4}{3}} z\big)\Big)K^{(1)}
-\frac{2}{3}\kappa^2\ell (\phi^{(0)})'(\phi^{(1)})'
&=& 
-\frac{\ell}{6}
  \Big(1-\frac{1}{2}\big(1+\Lambda\kappa^{\frac{4}{3}} z\big)\Big){}^{(h)}R (x^{\mu})
-\frac{\kappa^2\ell^2}{6}(\phi^{(0)})' {}^{(h)}\Box\varphi (x^{\mu})
\nonumber\\
&+&\frac{\kappa^2\ell}{6}D^{\rho}\varphi (x^{\mu})
                                    D_{\rho}\varphi (x^{\mu}). 
\eea
The first order part of the bulk scalar field equation of motion \eqref{scalar_eom} is given by 
\bea
\label{scalar_1st_eq}
\frac{1}{a(y)^4}\Big(a(y)^4 K{}^{(1)}\Big)_{,y}
-\frac{ 1} {3a(y)^2} {}^{(h)} R (x^{\mu})=0,
\eea
which is integrated as 
\bea
\label{scalar_sol}
K^{(1)}=-\frac{\ell}{6a(y)^2}{}^{(h)}R (x^{\mu})+\frac{\Phi(x^{\mu})}{a(y)^4},
\eea
where $\Phi(x^{\mu})$ is another integration constant which is also determined by the boundary condition on the brane.
We note that $\phi^{(1)}$ does not appear in the first order part of the bulk scalar field equation of motion \eqref{scalar_1st_eq},
because of the triviality of the bulk scalar field equation of motion at the zeroth order.

We then discuss the boundary conditions on the brane.
The first order part of the metric boundary condition \eqref{metric_z2} is given by
\bea
\label{1stmetric}
&-&\Big(1-\frac{1}{2}\big(1+\Lambda\kappa^{\frac{4}{3}} z\big)\Big)
\big(
 K^{(1)}{}^{\mu}{}_{\nu}
-K^{(1)}\delta{}^{\mu}{}_{\nu}
\big)
-\frac{1}{2}\kappa^2\ell (\phi^{(0)})'(\phi^{(1)})'\delta^{\mu}{}_\nu
\nonumber\\
&=&\frac{\kappa^2}{2}
\Big\{\tau^{\mu}{}_{\nu} (x^{\mu})
+\frac{1}{3}\ell^2(\phi^{(0)})'
 \big(D^{\mu}D_{\nu}\varphi (x^{\mu})
-\delta^\mu{}_\nu{}^{(h)}\Box\varphi (x^{\mu})
\big)
\nonumber\\
&+&\frac{2\ell}{3}
 \big(D^{\mu}\varphi (x^{\mu}) D_{\nu}\varphi (x^{\mu})
-\frac{1}{4}\delta^\mu{}_\nu D^\rho\varphi   (x^{\mu})D_\rho\varphi  (x^{\mu})\big)
\Big\}.
\eea
Noting that $U(\varphi)$ defined in \eqref{kim} is of $O(\epsilon)$,
the first order part of the scalar field boundary condition \eqref{scalar_z2} 
is given by
\bea
\label{1stscalar}
\Big(
K^{(1)}
+\frac{\ell}{6}{}^{(h)}R (x^{\mu})
\Big)
\big(\phi^{(0)}\big)'
+{}^{(h)}\Box\varphi (x^{\mu})
=-\frac{1}{\ell}U(\varphi).
\eea
Taking the trace of the first order part of the junction condition \eqref{1stmetric}
and combining it with the scalar field junction condition \eqref{1stscalar},
we obtain
\bea
\label{1st1}
&&\Big(1-\frac{3}{2}\big(1+\Lambda\kappa^{\frac{4}{3}} z\big)\Big)K^{(1)}
-\frac{2}{3}\kappa^2\ell^2(\phi^{(0)})'(\phi^{(1)})'
\nonumber\\
&=& \frac{\ell}{6} \big(1+\Lambda\kappa^{\frac{4}{3}} z\big){}^{(h)}R (x^{\mu})
+\frac{1}{6}\kappa^2\tau　 (x^{\mu})
-\frac{1}{3}\kappa^2\ell^2(\phi^{(0)})'
\Big({}^{(h)}\Box\varphi　 (x^{\mu})
+\frac{1}{2\ell}U(\varphi)
\Big).
\eea
Comparing Eq. (\ref{1st1}) with (\ref{1st2}),
we obtain
\bea
\label{trace}
\kappa^2\ell
\big(\phi^{(0)}\big)'
\Big(
{}^{(h)}\Box  \varphi (x^{\mu})
+\frac{1}{\ell}U(\varphi)
\Big)
-\Big(1+\frac{1}{2}\big(1+\Lambda\kappa^{\frac{4}{3}} z\big)\Big){}^{(h)}R (x^{\mu})
+\kappa^2 D^{\rho}\varphi (x^{\mu}) D_{\rho}\varphi (x^{\mu})
=\frac{\kappa^2}{\ell}\tau (x^{\mu}),
\eea
where the scalar field derivative with respect to $y$, $(\phi^{(1)})'$, does not appear.
The integration constant $C(x^{\mu})$ in Eq. (\ref{trace_sol}) is determined via
the trace of \eqref{1stmetric} and \eqref{trace} as
\bea
\label{cdet}
  C(x^{\mu})
=\frac{\kappa^2\ell^2}{6}
\big(\phi^{(0)}\big)'
\Big({}^{(h)}\Box \varphi (x^{\mu})
+\frac{1}{\ell}U(\varphi)
\Big).
\eea
Similarly, 
the integration constant $\Phi(x^{\mu})$ in Eq. (\ref{scalar_sol}) is determined via
\eqref{1stscalar} as
\bea
\label{phidet}
  \Phi(x^{\mu})
=-\frac{1}{\big(\phi^{(0)}\big)'}
\Big({}^{(h)}\Box \varphi (x^{\mu})
+\frac{1}{\ell}U(\varphi)
\Big).
\eea
With \eqref{cdet} and \eqref{phidet},
it is straightforward to check 
that the first order derivatives with respect to $y$,
\eqref{trace_sol} and  (\ref{scalar_sol}),
consistently
satisfy the constraint relation in the whole bulk \eqref{const_bulk}.

Eliminating $\Sigma^{(1)\mu}{}_{\nu}$ from \eqref{traceless} with the metric boundary condition \eqref{1stmetric},
we obtain
\bea
\label{traceless2}
\Big(1+\frac{1}{2}\big(1+\Lambda\kappa^{\frac{4}{3}} z\big)\Big)
\Big(
{}^{(h)}R^\mu{}_\nu (x^{\mu})
-\frac{1}{4}\delta^{\mu}{}_\nu{}^{(h)}R (x^{\mu})
\Big)
&=&\frac{\kappa^2}{\ell}
\Big(\tau^{\mu}{}_\nu (x^{\mu})
-\frac{1}{4}\delta^{\mu}{}_{\nu}\tau (x^{\mu})
\Big)
+\kappa^2
 \big(D^{\mu}\varphi  (x^{\mu}) D_{\nu}\varphi (x^{\mu})
\nonumber\\
&-&\frac{1}{4}\delta^\mu{}_\nu D^\rho\varphi  (x^{\mu})D_\rho\varphi (x^{\mu})
\big)
+\psi^{\mu}{}_{\nu} (x^{\mu}).
\eea
Combining the trace and traceless parts, Eqs. \eqref{trace} and \eqref{traceless2}, we obtain
\bea
\label{Einstein_eq}
\Big(1+\frac{1}{2}\big(1+\Lambda\kappa^{\frac{4}{3}} z\big)\Big)
{}^{(h)}G^\mu{}_\nu (x^{\mu})
&=&\frac{\kappa^2}{\ell}\tau^{\mu}{}_\nu (x^{\mu})
+\kappa^2
 \big(D^{\mu}\varphi (x^{\mu}) D_{\nu}\varphi (x^{\mu})
-\frac{1}{2}\delta^\mu{}_\nu D^\rho\varphi  (x^{\mu})D_\rho\varphi (x^{\mu})\big)
\nonumber\\
&-&\frac{\kappa^2\ell}{4} (\phi^{(0)})'
\Big({}^{(h)}\Box\varphi (x^{\mu})
+\frac{1}{\ell}U(\varphi)
\Big)\delta^{\mu}{}_{\nu}
+\psi^{\mu}{}_{\nu} (x^{\mu}).
\eea 
Eq. (\ref{Einstein_eq}) can be seen as the effective gravitational equation on the brane.
Thus the effective theory on the brane is given by the four-dimensional Einstein-scalar
theory with the corrections from the bulk,
where the effective four-dimensional gravitational constant 
and the four-dimensional scalar field are given by
\bea
\label{eff41}
\kappa_4^2:= \frac{1}{1+\frac{1}{2}\big(1+\Lambda\kappa^{\frac{4}{3}} z\big)}\frac{\kappa^2}{\ell},
\eea
and
\bea
\label{eff42}
\varphi_4(x^{\mu}):=\ell^{\frac{1}{2}}\varphi(x^{\mu}),
\eea
respectively.
In order to obtain the positive effective gravitational constant $\kappa_4^2>0$, 
we obtain the lower bound on $(1+\Lambda\kappa^{\frac{4}{3}} z)$.
Combining it with \eqref{condle}, we obtain the bound on 
the combination $(1+\Lambda z\kappa^{\frac{4}{3}} )$
\bea
\label{as}
-2<1+\Lambda\kappa^{\frac{4}{3}} z\leq 0.
\eea
The traceless component $\psi^{\mu}{}_{\nu}(x^{\mu})$ corresponds to the (generalized) dark radiation 
as in the Randall-Sundrum model \cite{ge1}.

We then investigate the energy-momentum conservation law on the brane.
The $O(\epsilon^{\frac{3}{2}})$ part of the $(y,\mu)$ component of the bulk gravitational equation \eqref{eq}
is given by 
\bea
\label{ymu}
0
&=&\Big(1-\frac{1}{2}\big(1+\Lambda\kappa^{\frac{4}{3}} z\big)\Big)
\big(
D_{\mu} K^{(1)}{}^{\mu}{}_{\nu}
-D_{\nu}K^{(1)}
\big)
-\frac{\kappa^2\ell }{2}(\phi^{(0)})'  K^{(1)}D_{\nu}\varphi (x^{\mu})
\nonumber\\
&+&\frac{\kappa^2\ell}{2} (\phi^{(0)})'   (D_{\nu}\phi^{(1)})'
+\frac{\kappa^2\ell^2}{6a(y)^2}
(\phi^{(0)})' {}^{(h)} G_{\nu}{}^{\mu}  (x^{\mu})D_{\mu} \varphi (x^{\mu})
\nonumber\\
&-&\frac{\kappa^2\ell}{6a(y)^2}
\Big( D_{\nu}\varphi  (x^{\mu}){}^{(h)}\Box\varphi (x^{\mu})
     -D_{\nu}D^{\mu}\varphi (x^{\mu}) D_{\mu}\varphi (x^{\mu})
\Big).
\eea
Substituting the metric and scalar field boundary conditions on the brane
Eqs. (\ref{1stmetric}) and (\ref{1stscalar}) into the brane limit of \eqref{ymu},
we obtain 
\bea
\label{mat}
D_{\mu}\tau^{\mu}{}_{\nu} (x^{\mu})
=U(\varphi) D_{\nu}\varphi (x^{\mu}).
\eea
Thus if there is no coupling of the brane matter to the bulk scalar field, $U(\varphi)=0$,
the energy-momentum tensor of the matter on the brane is conserved.
Then, taking the divergence of \eqref{Einstein_eq}
with the contracted Bianchi identities $D_{\mu}{}^{(h)}G^{\mu}{}_{\nu}=0$,
we obtain the generation of the dark radiation on the brane
\bea
\label{final}
D_\mu \psi^{\mu}{}_{\nu} (x^{\mu})
=\frac{\kappa^2\ell}{4} (\phi^{(0)})'
D_{\nu}
\Big({}^{(h)}\Box\varphi (x^{\mu})
+\frac{1}{\ell}U(\varphi)
\Big)
-\kappa^2
\Big({}^{(h)}\Box\varphi (x^{\mu})
+\frac{1}{\ell}U(\varphi)
\Big)
D_\nu\varphi (x^{\mu}).
\eea
Thus in the case that ${}^{(h)}\Box\varphi (x^{\mu})+\frac{1}{\ell}U(\varphi)=0$,
there is no generation of the dark radiation  by the bulk scalar field,
but then from \eqref{mat}
the energy-momentum tensor on the brane is not conserved.
On the other hand,
in the case that $U(\varphi)=0$,
where the energy-momentum tensor on the brane is conserved,
the dark radiation is still generated,
since the right-hand side of \eqref{final} is nonzero.
With Eqs. (\ref{beq1}), (\ref{trace_sol}), (\ref{cdet}), \eqref{phidet} and \eqref{final},
it is also straightforward to check that the constraint (\ref{ymu})
is satisfied in the whole bulk.

Integrating \eqref{beq1}, \eqref{trace_sol} and \eqref{scalar_sol}
with respect to $y$ once more,
with the use of the boundary conditions $q^{(1)}_{\mu\nu}(x^{\mu},0)=0$ and $\phi^{(1)}(x^\mu,0)=0$,
we find the solutions of the metric and the scalar field 
at the first order of the gradient expansion
\bea
\Big(1-\frac{1}{2}\big(1+\Lambda z\kappa^{\frac{4}{3}}\big)\Big)q^{(1)}_{\mu\nu}(x^{\mu},y)
&=&-\frac{\ell^2}{2}\Big(\frac{1}{a(y)^2}-1\Big)
\Big\{
\Big(1
+\frac{1}{2}\big(1+\Lambda z\kappa^{\frac{4}{3}}\big)
\Big)
\Big({}^{(h)}R_{\mu\nu}(x^{\mu})-\frac{1}{6}{}^{(h)}R(x^{\mu}) h_{\mu\nu} (x^{\mu})\Big)
\nonumber\\
&-&\frac{1}{12}\big(1+\Lambda z\kappa^{\frac{4}{3}}\big)
{}^{(h)}R (x^{\mu})
h_{\mu\nu}(x^{\mu})
\Big\}
\nonumber\\
&-&\frac{\kappa^2\ell^3}{6}(\phi^{(0)})'
\Big(\frac{1}{a(y)^2}-1\Big)
\Big(
D_{\mu}D_{\nu}\varphi(x^{\mu})
-\frac{1}{4}h_{\mu\nu}(x^{\mu}) {}^{(h)}\Box \varphi(x^{\mu})
\Big)
\nonumber\\
&+&\frac{\kappa^2\ell^2}{6}
\Big(\frac{1}{a(y)^2}-1\Big)
\Big(
D_{\mu}\varphi(x^{\mu})D_{\nu}\varphi(x^{\mu})
-\frac{1}{4}h_{\mu\nu}(x^{\mu})D^{\rho} \varphi(x^{\mu})D_{\rho} \varphi(x^{\mu})
\Big)
\nonumber\\
&+&\frac{\ell^2}{4}
\Big(\frac{1}{a(y)^4}-1\Big)
\Big\{
\psi_{\mu\nu}(x^{\mu})
\nonumber\\
&-&\frac{1}{2\ell (\phi^{(0)})'}
\Big(1
-\frac{1}{2}\big(1+\Lambda z\kappa^{\frac{4}{3}}\big)\Big)
\Big(
{}^{(h)}\Box  \varphi (x^{\mu})
+\frac{1}{\ell}U(\varphi) 
\Big)
h_{\mu\nu}(x^{\mu})
\Big\},
\eea\
and 
\bea
\kappa^2(\phi^{(0)})'\phi^{(1)}(x^{\mu},y)
&=&\frac{\ell}{8}
\Big(\frac{1}{a(y)^2}-1\Big)
\Big\{
\big(1+\Lambda z\kappa^{\frac{4}{3}}\big)
{}^{(h)}R(x^{\mu})
+\kappa^2 \ell (\phi^{(0)})'{}^{(h)}\Box\varphi(x^{\mu})
-\kappa^2D^{\rho}\varphi(x^{\mu})D_{\rho}\varphi(x^{\mu})
\Big\}
\nonumber\\
&-&\frac{3}{8}
\Big(\frac{1}{a(y)^4}-1\Big)
\frac{1}{(\phi^{(0)})'}
\Big(1-\frac{3}{2}\big(1+\Lambda z\kappa^{\frac{4}{3}}\big)\Big)
\Big(
{}^{(h)}\Box  \varphi (x^{\mu})
+\frac{1}{\ell}U(\varphi) 
\Big).
\eea


\section{Conclusions}
In this paper,
we have investigated the Randall-Sundrum type five-dimensional braneworld models,
focusing on the the roles of the field derivative coupling to the Einstein tensor in the bulk. 
The equations of motion remain of the second order.
We have first derived the generalized junction conditions
for the metric components and the scalar field.
The junction conditions contain the brane intrinsic quantities
because of the nonlinear terms in the equations of motion.

Next, with the use of these junction conditions,
we have derived the Minkowski and de Sitter braneworld solutions
which are embedded into the  $Z_2$-symmetric five-dimensional anti-de Sitter (AdS) spacetime.
We have derived the solutions of the AdS spacetime by solving the equations of motion in the bulk.
For the given metric ansatz of the five-dimensional AdS spacetime with 
the section of the four-dimensional Minkowski or de Sitter spacetime,
the scalar field equation of motion can be satisfied for the two cases:
In the first case, the AdS curvature scale has been related to the derivative coupling constant,
and the gravitational equations of motion have fixed the configuration of the scalar field.
In the second case, we could obtain the AdS solution with the section of the four-dimensional Minkowski spacetime
if the derivative coupling constant was tuned to the bulk cosmological constant.
In deriving the braneworld solutions,
we have focused on the first case which does not require the tuning relation
between the derivative coupling constant and the cosmological constant. 
We have found that the brane matter is not coupled to the bulk scalar field 
and the brane is supported by the constant brane tension
as in the Randall-Sundrum model.
The Minkowski brane solution can be obtained if the brane tension is tuned to the bulk contributions.
Once this tuning relation is violated, the de Sitter brane solutions have been obtained.
We have found two branches of the de Sitter brane solutions.
One has the smooth limit to the case with a trivial scalar field,
while the other does not.
For the same brane tension,
the brane expansion rate for the former branch is always greater than 
that of the latter branch.
As the brane tension increases,
for the former branch the brane expansion rate increases,
while for the latter branch this decreases.
Therefore, for the latter branch 
the expansion rate vanishes for the relatively larger value of the brane tension.

Finally, we have derived the low energy effective gravitational theory realized on the brane.
The low energy effective gravitational theory on the brane is given by the 
four-dimensional Einstein-scalar theory
with the corrections from the bulk.
If there is direct coupling of the brane matter to the bulk scalar field,
the energy-momentum tensor of the matter on the brane is not conserved.
The feature of the effective theory has given a bound on
the combination of the parameters in our model as \eqref{as}.

\section*{Acknowledgement}
This work was supported by the FCT-Portugal through the Grant No. SFRH/BPD/88299/2012.
We would like to thank the anonymous reviewer for his/her suggestions.
We also thank C. Germani for comments.

\appendix

\section{The tensor components}

\subsection{The general case}

In this subsection, we show the components of the curvature tensor 
for the general bulk spacetime described by the general Gaussian-normal coordinate \eqref{GN}. 

The nonzero components of the Christoffel symbol are given by
\bea
\Gamma^{y}_{\mu\nu}=-K_{\mu\nu},\quad
\Gamma^{\mu}_{\alpha\beta}={}^{(q)}\Gamma^{\mu}_{\alpha\beta},\quad
\Gamma^{\mu}_{y\alpha}=K^{\mu}{}_\alpha,
\eea
where the extrinsic curvature tensor on a constant-$y$ hypersurface is defined in \eqref{extrinsic}.

The nonzero components of the Riemann tensor are given by 
\bea
R^y{}_{\alpha y \beta}&=&-K_{\alpha\beta,y}+K_{\alpha\gamma}K^{\gamma}{}_{\beta},
\nonumber\\
R^{\alpha}{}_{y\beta y}
&=&-K^{\alpha}{}_{\beta,y}- K^{\alpha}{}_{\gamma}K^{\gamma}{}_\beta,
\nonumber\\
R^{\alpha}{}_{\mu\beta \nu}
&=&
{}^{(q)}R^{\alpha}{}_{\mu\beta \nu}
+K^{\alpha}{}_{\nu}K_{\mu\beta}
-K^{\alpha}{}_{\beta}K_{\mu\nu},
\nonumber\\
R^y{}_{\alpha\beta\gamma}&=&
-K_{\alpha\gamma|\beta}
+K_{\alpha\beta|\gamma},
\nonumber\\
R^{\alpha}{}_{y\mu\nu}&=&
 K^{\alpha}{}_{\nu|\mu}
- K^{\alpha}{}_{\mu|\nu}.
\eea

The nonzero components of the Ricci tensor are given by 
\bea
R_{yy}&=&-K_{,y} -K^{\alpha\beta}K_{\alpha\beta},
\nonumber\\
R_{\alpha\beta}
&=&-K_{\alpha\beta,y}
+{}^{(q)}R_{\alpha\beta}
+2K_{\alpha\gamma} K^{\gamma}{}_{\beta}
-KK_{\alpha\beta},
\nonumber\\
R_{y\mu}
&=&K^{\alpha}{}_{\mu|\alpha}
    -K_{|\mu}.
\eea

The Ricci scalar is given by 
\bea
R=-2K_{,y} +{}^{(q)}R-K^{\alpha\beta}K_{\alpha\beta}-K^2.
\eea

The nonzero components of the Einstein tensor are given by 
\bea
\label{gyy}
G_{yy}&=&-\frac{1}{2}{}^{(q)}R-\frac{1}{2}K^{\alpha\beta}K_{\alpha\beta}
+\frac{1}{2}K^2,
\nonumber\\
G_{\alpha\beta}
&=&-K_{\alpha\beta,y}
+{}^{(q)}G_{\alpha\beta}
+2K_{\alpha\gamma}K^{\gamma}{}_\beta
-KK_{\alpha\beta}
+\frac{1}{2}q_{\alpha\beta}
\big(K^2+K^{\rho\sigma}K_{\rho\sigma}\big)
+q_{\alpha\beta}K_{,y},
\nonumber \\
G_{y\mu}
&=&K^{\alpha}{}_{\mu|\alpha}
    -K_{|\mu}.
\eea

The second order covariant derivatives of the scalar field are given by 
\bea
&&\nabla_y \nabla_y\phi= \phi'',
\quad
\nabla_\alpha\nabla_\beta\phi
=\phi_{|\alpha\beta}
+K_{\alpha\beta}\phi',
\nonumber\\
&&\nabla_y\nabla_\beta \phi
= \big(\phi_{|\alpha}\big)'
 -K_\alpha{}^\beta \phi_{|\beta},
\quad
\Box\phi
=\phi'' +K\phi' +{}^{(q)}\Box\phi.
\eea

\subsection{The warped five-dimensional spacetime}

In this subsection, we focus on the case of $D=5$ and the metric given by \eqref{gingin},
where the four-dimensional section of the metric is given by the maximally symmetric spacetime
with the scalar curvature $R[\gamma]=12 H^2$.

The nonzero components of the Christoffel symbol are given by 
\bea
\Gamma^{y}_{\mu\nu}=-\frac{a'}{a}q_{\mu\nu},\quad
\Gamma^{\mu}_{\alpha\beta}={}^{(q)}\Gamma^{\mu}_{\alpha\beta},\quad
\Gamma^{\mu}_{y\alpha}=\frac{a'}{a} \delta^{\mu}{}_\alpha,
\eea
where $a(y)$ is the warp factor.

The nonzero components of the Riemann tensor are given by 
\bea
R^y{}_{\alpha y\beta}&=&-\frac{a''}{a}q_{\alpha\beta},
\nonumber\\
R^\alpha{}_{y\beta y}&=&-\frac{a''}{a}\delta^{\alpha}{}_\beta,
\nonumber\\
R^\alpha{}_{\mu\beta\nu}
&=&\Big[
\Big(\frac{a'}{a}\Big)^2
-\frac{H^2}{a^2}
\Big]
\Big[
\delta^{\alpha}{}_{\nu}
q_{\mu\beta}
-\delta^{\alpha}{}_{\beta}
q_{\mu\nu}
\Big].
\eea

The nonzero components of the Ricci tensor are given by 
\bea
R_{yy}&=&-4\frac{a''}{a},
\nonumber\\
R_{\alpha\beta}&=&
\Big[-\frac{a''}{a}
+3\Big(\frac{H^2}{a^2}-\Big(\frac{a'}{a}\Big)^2\Big)
\Big]
q_{\alpha\beta}.
\eea

The Ricci scalar is given by 
\bea
R=-8\frac{a''}{a}
   +12\frac{H^2}{a^2}
    -12\Big(\frac{a'}{a}\Big)^2.
\eea

The nonzero components of the Einstein tensor are given by 
\bea
\label{k1}
G_{yy}&=&-6\frac{H^2}{a^2}+6\Big(\frac{a'}{a}\Big)^2,
\nonumber\\
G_{\alpha\beta}&=&
\Big[
  3\frac{a''}{a}
+3\Big(\frac{a'}{a}\Big)^2 
-3\frac{H^2}{a^2}
\Big]q_{\alpha\beta}.
\eea

Concerning the contributions from the nonminimal derivative coupling
\bea
\label{k2}
L_{yy}&=&-3\frac{H^2}{a^2}(\phi')^2
+9\Big(\frac{a'}{a}\Big)^2(\phi')^2,
\nonumber\\
L_{\alpha\beta}
&=&
\Big[3\frac{a'}{a}\phi'\phi''
+\frac{3}{2}\frac{a''}{a}(\phi')^2
+\frac{3}{2}\Big(\frac{a'}{a}\Big)^2(\phi')^2
+\frac{3}{2}\frac{H^2}{a^2}(\phi')^2
\Big]
q_{\alpha\beta}.
\eea

The nonzero components of the scalar field energy-momentum tensor  are given by
\bea
\label{k3}
T_{yy}&=&\frac{1}{2}(\phi')^2-\frac{\Lambda}{\kappa^2},
\nonumber\\
T_{\alpha\beta}&=&
-\frac{1}{2}(\phi')^2q_{\alpha\beta}-\frac{\Lambda}{\kappa^2} q_{\alpha\beta},
\eea
for the choice of $V(\phi)=\frac{\Lambda}{\kappa^2}$.

The scalar field equation of motion is given by 
\bea
\label{k4}
\Big[1-6 z\kappa^{\frac{4}{3}}\Big\{\Big(\frac{a'}{a}\Big)^2-\frac{H^2}{a^2}\Big\}\Big]\phi''
+\frac{4a'}{a}\phi'
\Big[
1-z\kappa^{\frac{4}{3}}
\Big(3\frac{a''}{a}+3\Big(\frac{a'}{a}\Big)^2-3\frac{H^2}{a^2}\Big)
\Big]
=0.
\eea

\end{document}